\documentclass[12pt,aaspp4]{aastex}

\usepackage{natbib}
\begin{document}
\newcommand{\up}[1]{\ifmmode^{\rm #1}\else$^{\rm #1}$\fi}
\newcommand{\zdot}{\makebox[0pt][l]{.}}
\newcommand{\upd}{\up{d}}
\newcommand{\uph}{\up{h}}
\newcommand{\upm}{\up{m}}
\newcommand{\ups}{\up{s}}
\newcommand{\arcd}{\ifmmode^{\circ}\else$^{\circ}$\fi}
\newcommand{\arcm}{\ifmmode{'}\else$'$\fi}
\newcommand{\arcs}{\ifmmode{''}\else$''$\fi}

\title{The Araucaria Project. The Distance to the Small Magellanic Cloud from Near-Infrared Photometry 
of RR Lyrae Variables 
\footnote{Based on observations obtained with the ESO NTT for programmes 082.D-0513(A) and 079.D-0482(A)} }
\author{Olaf Szewczyk}
\affil{Universidad de Concepci{\'o}n, Departamento de Astronom{\'i}a, Casilla 160-C, Concepci{\'o}n, Chile}
\authoremail{szewczyk@astro-udec.cl}
\author{Grzegorz Pietrzy{\'n}ski}
\affil{Universidad de Concepci{\'o}n, Departamento de Astronom{\'i}a, Casilla 160-C, Concepci{\'o}n, Chile}
\affil{Warsaw University Observatory, Al. Ujazdowskie 4, 00-478, Warsaw, Poland}
\authoremail{pietrzyn@astrouw.edu.pl}
\author{Wolfgang Gieren}
\affil{Universidad de Concepci{\'o}n, Departamento de Astronom{\'i}a, Casilla 160-C, Concepci{\'o}n, Chile}
\authoremail{wgieren@astro-udec.cl}
\author{Anna Ciechanowska}
\affil{Warsaw University Observatory, Al. Ujazdowskie 4, 00-478, Warsaw, Poland}
\authoremail{aciechan@astrouw.edu.pl}
\author{Fabio Bresolin}
\affil{Institute for Astronomy, University of Hawaii at Manoa, 2680 Woodlawn Drive, Honolulu HI 96822, USA}
\authoremail{bresolin@ifa.hawaii.edu}
\author{Rolf-Peter Kudritzki}
\affil{Institute for Astronomy, University of Hawaii at Manoa, 2680 Woodlawn Drive, Honolulu HI 96822, USA}
\authoremail{kud@ifa.hawaii.edu}

\begin{abstract}
We have obtained deep infrared $J$ and $K$ band observations of nine $4.9 \times 4.9$ arcmin  
fields in the Small Magellanic Cloud (SMC) with the ESO
New Technology Telescope equipped with the SOFI infrared camera. In
these fields, 34 RR Lyrae stars catalogued by the OGLE collaboration
were identified. Using different theoretical and empirical calibrations of the infrared
period-luminosity-metallicity relation, we find consistent SMC
distance moduli, and find a best true
distance modulus to the SMC of $18.97 \pm 0.03$ (statistical) $\pm$ 
0.12 (systematic) mag which agrees well with most independent distance
determinations to this galaxy, and puts the SMC 0.39 mag more distant
than the LMC for which our group has recently derived, from the same technique, a distance
of 18.58 mag.
\end{abstract}


\section{Introduction}
In our ongoing Araucaria Project (e.g. Gieren et al. 2005a), we are applying a
number of different stellar standard candles to independently determine the
distances to a sample of nearby galaxies.  The systematic differences
between the distance results obtained for the individual galaxies from
the various stellar candles will be analyzed in forthcoming
papers. This analysis is expected to lead to a detailed
understanding of how the various stellar techniques, which are
fundamental to calibrate the first rungs of the distance ladder, depend
on metallicity and age.  While the objects we use for the
distance determinations are usually detected from optical wide-field imaging
surveys of the target galaxies (e.g. Pietrzy{\'n}ski et al. 2002b), the most
accurate distance work is then done from follow-up near-infrared images which
virtually eliminate reddening as a significant source of error on the
results. Examples of this very successful approach are the Cepheid work on
the Sculptor galaxies NGC 300 (Gieren et al. 2005b),
NGC 55 (Gieren et al. 2008) and NGC 247 (Gieren et al. 2009), and the red clump star work 
on the LMC (Pietrzy{\'n}ski \& Gieren 2002). The Araucaria Project has also been
developing a new spectroscopic distance indicator, viz. the Flux-weighted Gravity-
Luminosity Relationship (FGLR) for blue supergiants (Kudritzki et al. 2003;
2008) which is able to yield distances accurate to 5\% to galaxies
containing massive blue stars out to about 10 Mpc from low-resolution spectra
(Kudritzki et al. 2008, Urbaneja et al. 2008).

Thanks to several recent theoretical and empirical studies, evidence has been
mounting that RR Lyrae (RRL) stars are excellent standard candles in the
near-infrared spectral range, providing distance results which are superior to
the traditional optical method (e.g. Bono 2003a).  Longmore et al. (1986) were
the first to show that RRL variable stars follow a period-luminosity (PL)
relation in the near-infrared K-band.  Their pioneering work was followed by
Liu \& Janes (1990), 
Jones et al. (1996),
and Skillen et al. (1993),
who applied infrared versions of the Baade-Wesselink method to
calibrate the luminosities and distances of RRL stars.  A very
comprehensive analysis of the IR properties of RRL stars was
given by Nemec et al. (1994).  The first theoretical constraints on the
K-band PL relation of RRL stars are based on non-linear convective
pulsation models which were presented by Bono et al. (2001).  Dall'Ora et
al. (2004) later demonstrated that the K-band PL relation for RRL
stars appears to have a very small scatter for globular clusters, with 
small intrinsic spread in metallicity for this type of stars. Further theoretical
explorations of the RRL period-mean magnitude-metallicity
relations in near-infrared passbands were carried out by Bono et
al. (2003b), Catelan et al. (2004), and Cassisi et al. (2004). Most recently, Sollima et
al. (2008) analyzed near-infrared K-band data of RRL stars in
some 15 Galactic globular clusters and provided the first empirical
calibration of the period-luminosity-metallicity (PLZ) relation in the
$K$ band. All these existing theoretical and empirical studies have
suggested that the RRL star K-band PLZ relation appears to be
a superb means to determine accurate distances to galaxies
hosting an abundant old stellar population. 

We have therefore started to include this tool in the
Araucaria Project distance work.  In previous papers, we determined the distance
to the Sculptor dwarf galaxy (Pietrzy{\'n}ski et al. 2008), and to the LMC
(Szewczyk et al. 2008) from this method. In this paper,
we apply the technique to a sample of RR Lyrae stars distributed over the SMC,
and determine the distance to the SMC for the first time with the infrared
RR Lyrae technique.

\section{Observations, Data Reduction and Calibration}

During several observing runs of the Araucaria Project, we have obtained all the data 
analyzed and presented in this paper with the SOFI infrared camera attached to the ESO
New Technology Telescope. The Large Field setup was used to yielding a 
$4.9 \times 4.9$ arcmin field of view and a scale of $0.288$ arcsec per pixel.

It was possible to collect deep $Js$ and $Ks$ images of six distinct SMC fields during 3 nights 
in 2007, and of three more fields over 2 nights in 2008. All but one nights were found 
to have photometric conditions. Detailed information on each observed field can be found 
in Table \ref{tabfields}. Figure \ref{figfields} illustrates the location of the observed 
fields on an image of the SMC. In order to take into account the rapid sky level 
variations in the IR passbands, we have been using a dithering technique during 
the observations.
 Total integration times per field were 60 min in the $Ks$ band, and 15 min in the $Js$ band.

For all the reductions and calibrations, the pipeline developed in the course of the Araucaria
 Project was used. As a first step, the subtraction of sky level subprocess (including the
  masking of stars with the IRAF xdimsum package) was applied (Pietrzy{\'n}ski \& Gieren 
  2002). Later, each single image was flatfielded and stacked into the final deep field. 
  PSF (point spread-function) photometry with aperture corrections was then performed 
  in the same way as described in Pietrzy{\'n}ski et al. (2002b).

The photometry was calibrated onto the standard system using observations of 
15 standard stars from the UKIRT list (Hawarden et al. 2001). All of them were observed 
under photometric conditions at different airmasses spread in between the regular target fields 
acquisition. The data obtained during the non-photometric night was compared with nearby 
photometric data and the zero-point difference was estimated. Thanks to the large number of 
standard stars observed, the accuracy of our photometric zero points was estimated to be 
as good as $0.02$ mag. The results were compared with the Two Micron All Sky Survey (2MASS)
 catalog for common stars. This way we could determine the zero point differences which are 
 given in Table \ref{tab2mass}. The calibrated NIR magnitudes for all RRL stars identified 
 in our science target fields are presented in Table \ref{taballstars}. All uncertainties 
 given in the Tables are standard deviations.

\section{Near-Infrared Period-Luminosity Relations}

The final sample of 34 RRL stars analyzed in our SOFI/NTT fields was chosen by rejecting 
all stars which were blended with or contaminated by other close-by objects in the frames. 
Then it was cross-identified with the Optical Gravitational Lensing Experiment (OGLE) catalog
of RRL stars in the SMC (Soszy{\'n}ski et al. 2002). The positions of the RRL stars in the 
$K$, $J-K$ color-magnitude diagram are shown in Figure \ref{figcmd}. All variables have at least 
one measurement collected during photometric conditions. In the case of RRL stars which were 
observed more than once, we took a straight average of the random-phase magnitudes, which 
is expected to lead to a better approximation of their mean magnitudes. 

Following the identification from the OGLE catalog (Soszy{\'n}ski et al. 2002),
 we fundamentalize the RRc periods by adding $log(P)=0.127$ in
 order to combine the RRab and RRc stars. 
 We have corrected the observed magnitudes for extinction, adopting
 the reddening values published by Udalski et al. (1999) which are shown in Table \ref{tabfields}. 
 Adopting the reddening law from Schlegel et al (1998), we calculate the following selective 
 extinctions in the different bands: $A_{K} = 0.367E(B-V)$ and $A_{J}= 0.902E(B-V)$. The 
 final photometric data for the sample of RRL stars are presented in Table \ref{tabrrlyr}.
The PL relations for the $J$ and $K$ bands derived from our magnitudes
are shown in Figure \ref{figpmd}. 

The relatively large scatter seen in Figures \ref{figcmd} and \ref{figpmd} are mainly 
caused by four factors: (1) the random single-phase nature of our IR measurements, which 
represents the mean magnitude of an RRL variable only to $\sim0.15$ mag (e.g. Del Principe 
et al. 2006), (2) the metallicity spread among the RRL stars in SMC, (3) the depth extension
of the SMC in the line of sight,
and (4) the accuracy of our single measurements which is $0.03-0.11$ mag for stars of 
magnitudes of $18.0-19.0$ mag in the $K$-band.

We compare the PL relations derived from the observed RRL stars against the existing 
theoretical (Bono et al. 2003; Catelan et al. 2004) and empirical (Sollima et al. 2008)
relations. Table \ref{tabpl} lists our results for the slope and zero-point values 
of the PL relations and as well as those from the theoretical and empirical ones.

\section{The Distance Determination}

In order to derive the apparent distance moduli to the SMC from our data, we used the 
following calibratons of the NIR PL relations of mixed population RRL stars:

\begin{equation} M_{K} = -1.07 - 2.38\log P + 0.08[Fe/H]
\qquad\mbox{-- Sollima et al. (2008)}
\end{equation}

\begin{equation} M_{K} = -0.77 - 2.101\log P + 0.231[Fe/H]
\qquad\mbox{-- Bono et al. (2003b)}
\end{equation}

\begin{equation} M_{K} = -0.597 - 2.353\log P + 0.175\log Z
\qquad\mbox{-- Catelan et al. (2004)}
\end{equation}

\begin{equation} M_{J} = -0.141 - 1.773\log P + 0.190\log Z
\qquad\mbox{-- Catelan et al. (2004)}
\end{equation}

We recall that the calibration of Sollima et al. (2008) was constructed for the 2MASS photometric 
system, while the calibrations of Catelan et al. (2004) and Bono et al. (2003) are usable 
for the Glass or Bessel \& Brett systems, respectively. Therefore we transformed our own 
data, calibrated onto the UKIRT system (Hawarden et al. 2001), to the 
Glass and Bessel \& Brett systems using the transformations presented by Carpenter (2001) 
before calculating distances using the calibrations of Catelan et al. (2004) and 
Bono et al. (2003). Since there is no signifficant difference between the $K$-band of 
2MASS and UKIRT systems (Carpenter 2001), we did not apply any transformations to our data 
while using the Sollima et al. (2008) calibration.

Assuming the mean metallicity of RRL stars in the SMC to be $[Fe/H] = -1.7$ dex 
(e.g. Udalski 2000) we have calculated the $K$- and $J$-band distance moduli for our sample 
of RRL stars using the final photometric magnitudes as given in Table \ref{tabrrlyr}. The 
fits for the relations $(1)-(4)$ to both sets of data are displayed in 
Figure \ref{figpmdfit}.

The true distance moduli of the SMC obtained from the different PLZ
calibrations are summarized in Table \ref{tabdist}. As our best distance determination, 
we adopt the average value from the different calibrations, yielding $18.97$ mag. The
uncertainty on this value will be discussed in the next section.

\section{Discussion}

The distance moduli obtained from several independent theoretical
and empirical calibrations of the infrared RR Lyrae PLZ relation
 are consistent. The maximum difference 
of 0.06 mag between the results from the calibrations of Sollima et al.
(2008) and Bono et al. (2003b) is certainly not significant taking into 
account all the uncertainties, which affect the whole process of
constructing these calibrations. It is interesting to note 
that a very similar difference between the distance moduli derived using
these two calibrations was recently obtained by Pietrzy{\'n}ski et al. 
(2008) for the Sculptor galaxy ($[Fe/H] = -1.83$ dex). Therefore,
perhaps there is just a zero point offset in the sense that the
distances from the  calibration of Sollima et al. (2008) are slightly
shorter compared to those from the calibration of Bono et al. (2003b).

Taking into account the errors associated with the adopted calibrations,
mean metallicity, photometric zero point and absorption correction,
we estimate the systematic error of our distance determination to be of
0.12 mag. Therefore our best distance modulus determination to the SMC
is: $18.97 \pm 0.03$ (statistical) $\pm 0.12$ (systematic) mag. This
value agrees very well with the value of $18.967 \pm 0.018$ mag derived
from K-band photometry of red clump stars (Pietrzy{\'n}ski, Gieren \& Udalski 2003).
It also agrees, within the combined $1\sigma$ uncertainties, with
 the true SMC distance modulus derived from Cepheid variables (e.g.
Udalski 2000; Barnes et al. 1993). The
distance modulus of the SMC found in this paper is 0.39 mag larger
than the value for the LMC we found from the same technique (Szewczyk et al. 2008).
Again, this difference is very much in line with modern values 
found from other techniques, like Cepheid variables (0.5 mag; Udalski 2000),
and red clump stars (0.47 mag; Pietrzy{\'n}ski, Gieren \& Udalski 2003).

We do not observe any significant difference in the dispersions of the NIR P-L
relations defined by the RR Lyrae variables in the SMC (this paper), and the LMC
(Szewczyk et al. 2008), arguing against a more significant distance spread
in the line of sight in the SMC than in the SMC, as usually found for younger
tracers like Cepheids.

\section{Summary and Conclusions}

The results of our deep infrared imaging and photometry of 34 RR Lyrae stars 
spread over nine $4.9 \times 4.9$ arcmin fields 
in the SMC are presented.  Our data show two clear sequences
in the period luminosity plane, which correspond to the RRc and RRab
stars. After fundamentalizing the RRc periods to the period of
the RRab stars by adding $\log P = 0.127$, both groups were
merged, and the distance moduli to the SMC were determined using
different theoretical and empirical calibrations.  Our final adopted
distance agrees very well with modern results obtained from 
a number of different techniques. 

Our results confirm that the RR Lyrae period-luminosity-metallicity 
relations in the near-infrared passbands are potentially a very good tool 
for accurate distance measurements. In future programs, particularly 
for the LMC, we want to reduce the systematic uncertainty of our
distance determinations with this method by providing multi-phase JK observations 
of the RR Lyrae variables to determine their mean magnitudes
with better precision.

\acknowledgements

WG and  GP  gratefully acknowledge financial support for this
work from the Chilean Center for Astrophysics FONDAP 15010003,
and from the BASAL Center for Astrophysics and Related Tecnologies (CATA 
2007 PFB 06). Support from the Polish grant N203 002 31/046 and the FOCUS
subsidy of the Fundation for Polish Science (FNP)
is also acknowledged.
It is a special pleasure to thank the support astronomers at ESO-La Silla 
for their expert help in the observations, and the ESO OPC for the
generous amounts of observing time at the NTT allocated to our programme.

\setcounter{figure}{0}

\clearpage
\begin{figure}[p]
\includegraphics{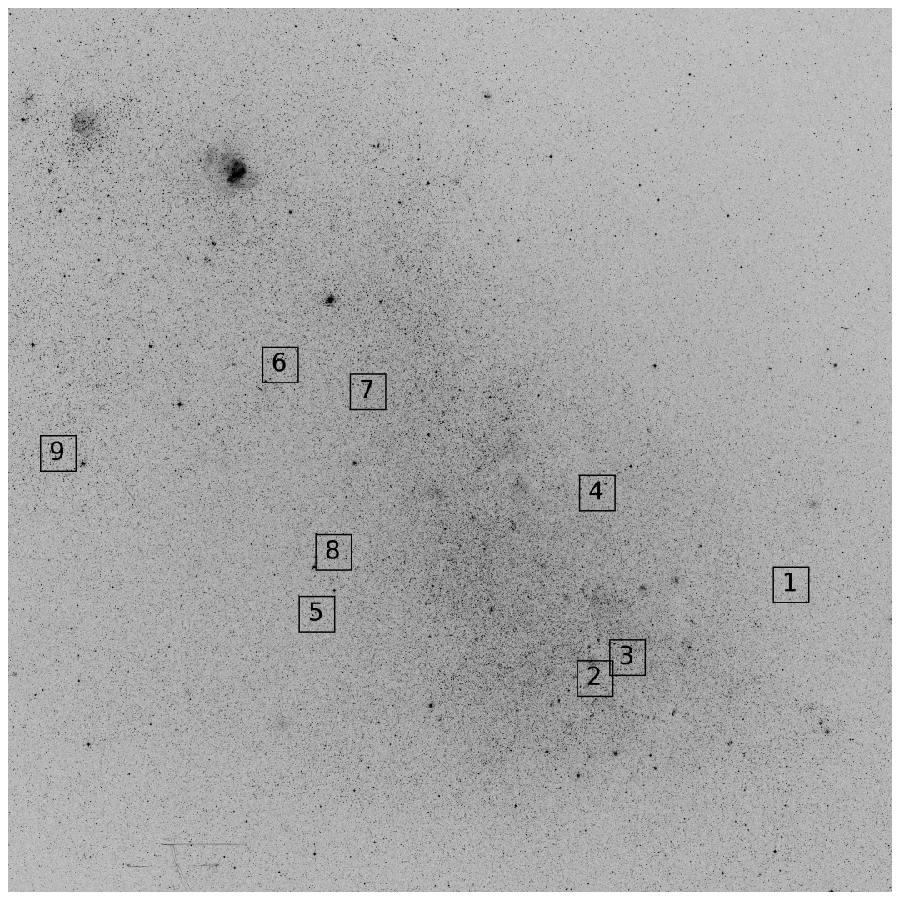}
\caption{The location of our observed $4.9 \times 4.9$ arcmin NTT/SOFI fields in the SMC 
on the DSS-1 plate. North is up and east to the left.}
\label{figfields}
\end{figure}

\clearpage
\begin{figure}[p]
\includegraphics{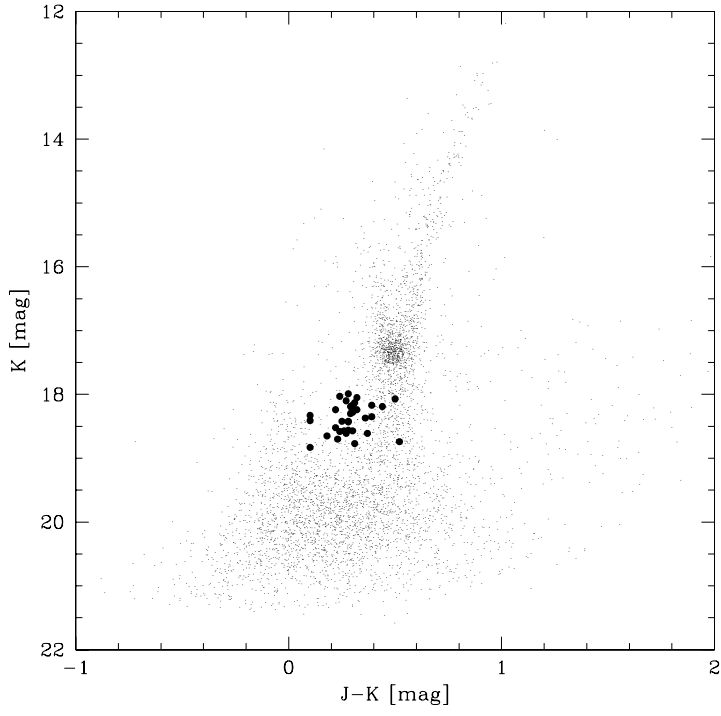}
\caption{The infrared color-magnitude diagram, showing the locations of the
RRL stars we have identified in the observed fields.}
\label{figcmd}
\end{figure}

\clearpage
\begin{figure}[p]
\includegraphics{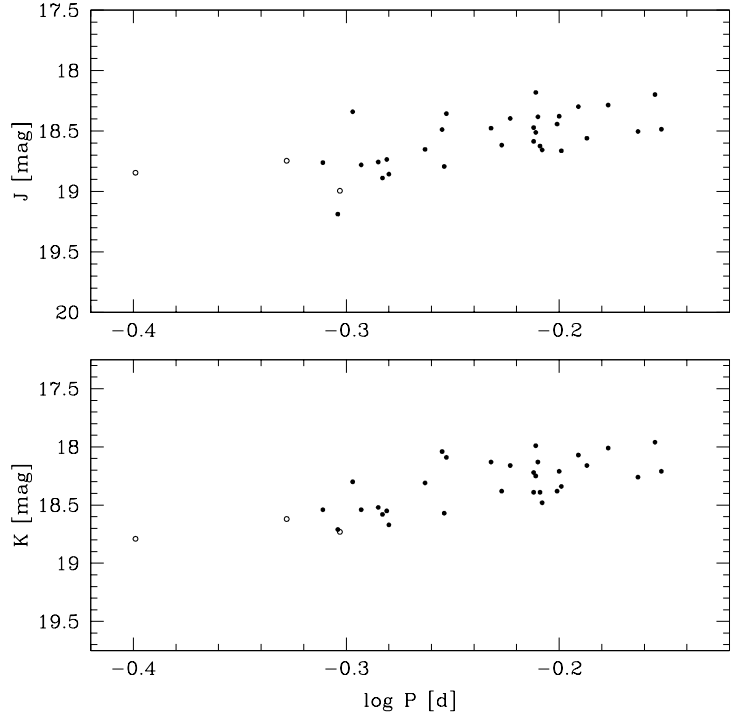}
\caption{The near-infrared $K$ and $J$ band period-luminosity relations defined by the RR Lyrae 
stars observed in the SMC. Period is in days. Fundamental mode pulsators are indicated
with filled circles, first overtone pulsators with open circles.}
\label{figpmd}
\end{figure}

\clearpage
\begin{figure}[p]
\includegraphics{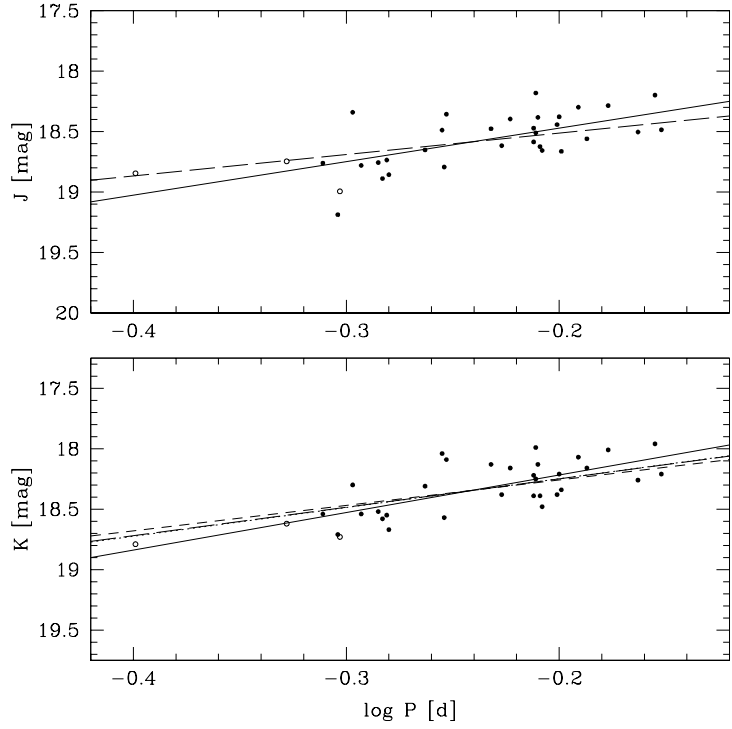}
\caption{The PL relations in $K$ and $J$ defined by the RR Lyrae stars observed in the SMC, plotted 
along fith the best fitting lines. The solid, dotted, short- and long-dashed lines correspond 
to the free fit, and the calibrations of Sollima et al. (2008), Bono et al. (2003b) and 
Catelan et al. (2004), respectively.}
\label{figpmdfit}
\end{figure}

\setcounter{table}{0}

\clearpage
\begin{deluxetable}{ccccccccc}
\tablewidth{0pc}
\tabletypesize{\tiny}
\tablecaption{Observational information on the target fields.
Extinction values are taken from the reddening maps of Udalski et al. (1999).}
\tablehead{ \colhead {Field} & \colhead{Field} & \colhead {RA2000} & \colhead {DEC2000} & \colhead {Date of} & \colhead {MJD of} & \colhead {MJD of} & \colhead {Conditions} & \colhead {Extinction} \\
\colhead {No} & \colhead {name} & \colhead {} & \colhead {} & \colhead {observation} & \colhead {$Js$ exposure} & \colhead {$Ks$ exposure} & \colhead {} & \colhead {E(B-V)} }
\startdata
1 & SC2\_FI & 00:41:57.60 & -73:04:30.00 & 2007-08-27 & $54340.034307$ & $54340.045077$ & CLR & $0.078$ \\
&  &  &  & 2007-08-28 & $54341.170250$ & $54341.178630$ & STD &  \\
2 & SC4\_FII & 00:47:55.20 & -73:18:36.00 & 2007-08-27 & $54340.230586$ & $54340.241368$ & CLR & $0.094$ \\
&  &  &  & 2007-08-28 & $54341.373742$ & $54341.382950$ & STD &  \\
3 & SC4\_FIII & 00:46:55.20 & -73:15:36.00 & 2007-08-27 & $54340.152757$ & $54340.182271$ & CLR & $0.094$ \\
&  &  &  & 2007-08-28 & $54341.348700$ & $54341.363482$ & STD &  \\
4 & SC4\_FIV & 00:48:04.80 & -72:53:24.00 & 2007-08-27 & $54340.289348$ & $54340.289348$ & CLR & $0.094$ \\
&  &  &  & 2007-11-24 & $54429.106746$ & $54429.117547$ & STD &  \\
5 & SC7\_FV & 00:56:43.20 & -73:10:30.00 & 2007-08-28 & $54341.195314$ & $54341.206117$ & STD & $0.097$ \\
6 & SC8\_FVI & 00:57:50.40 & -72:36:36.00 & 2007-08-28 & $54341.276830$ & $54341.276830$ & STD & $0.100$ \\
7 & SC7\_FVII & 00:55:09.15 & -72:40:12.80 & 2008-12-14 & $54815.030191$ & $54815.041781$ & STD & $0.097$ \\
8 & SC7\_FVIII & 00:56:11.44 & -73:02:08.70 & 2008-12-14 & $54815.091332$ & $54815.102922$ & STD & $0.097$ \\
9 & SC10\_FIX & 01:04:38.40 & -72:48:00.00 & 2008-12-15 & $54816.038947$ & $54816.050542$ & STD & $0.079$
\enddata
\label{tabfields}
\end{deluxetable}

\begin{deluxetable}{ccc}
\tablewidth{0pc}
\tabletypesize{\small}
\tablecaption{Difference of zero point estimation between 2MASS and selected observational data.}
\tablehead{\colhead{Field}&\colhead{$|$2MASS$-Js|$}&\colhead{$|$2MASS$-Ks|$} \\
\colhead{name}&\colhead{[mag]}&\colhead{[mag]}}
\startdata
SC7\_FV & $0.06\pm0.13$ & $0.06\pm0.08$ \\
SC7\_FVIII & $0.07\pm0.04$ & $0.05\pm0.13$
\enddata
\label{tab2mass}
\end{deluxetable}

\begin{deluxetable}{ccccccc}
\tablecaption{Individual $Js$ and $Ks$ Band Observations of RR Lyrae stars in SMC fields}
\tablewidth{0pc}
\tabletypesize{\footnotesize}
\tablehead{\colhead{Star ID} & \colhead{Star} & \colhead{Period} & \colhead{Field} & \colhead{$Js$} & \colhead{$Ks$} & \colhead{Remarks} \\
\colhead{[OGLE]} & \colhead{type} & \colhead{[d]} & \colhead{name} & \colhead{[mag]} & \colhead{[mag]} & \colhead {}}
\startdata
\multicolumn{7}{c}{2008-08-27}\\
\hline
OGLE004211.68-730329.2 & ab & $0.632047$ & SC2\_FI & $18.724\pm0.047$ & $18.302\pm0.043$ &  \\
OGLE004225.29-730349.8 & ab & $0.699868$ & SC2\_FI & $18.341\pm0.039$ & $18.088\pm0.039$ &  \\
OGLE004126.84-730355.2 & ab & $0.705041$ & SC2\_FI & $18.673\pm0.047$ & $18.217\pm0.043$ &  \\
OGLE004154.36-730539.3 & ab & $0.524564$ & SC2\_FI & $18.850\pm0.045$ & $18.558\pm0.041$ &  \\
OGLE004156.46-730641.7 & ab & $0.586298$ & SC2\_FI & $17.568\pm0.027$ & $17.111\pm0.034$ & blend \\
OGLE004817.88-731815.2 & ab & $0.558981$ & SC4\_FII & $18.441\pm0.039$ & $18.232\pm0.038$ &  \\
OGLE004744.69-731919.1 & ab & $0.598960$ & SC4\_FII & $18.491\pm0.052$ & $18.193\pm0.038$ &  \\
OGLE004805.65-732033.9 & ab & $0.523798$ & SC4\_FII & $18.748\pm0.039$ & $18.479\pm0.036$ &  \\
OGLE004724.96-731427.2 & ab & $0.614266$ & SC4\_FIII & $18.671\pm0.058$ & $18.419\pm0.040$ &  \\
OGLE004649.00-731544.5 & ab & $0.665961$ & SC4\_FIII & $18.383\pm0.041$ & $18.126\pm0.031$ &  \\
OGLE004650.36-731652.2 & ab & $0.630375$ & SC4\_FIII & $18.495\pm0.033$ & $18.244\pm0.022$ &  \\
OGLE004755.07-725141.5 & ab & $0.504549$ & SC4\_FIV & $18.395\pm0.042$ & $18.302\pm0.034$ &  \\
OGLE004805.17-725144.4 & ab & $0.509540$ & SC4\_FIV & $18.728\pm0.054$ & $18.515\pm0.050$ &  \\
OGLE004754.45-725212.1 & ab & $0.586127$ & SC4\_FIV & $18.475\pm0.070$ & $18.167\pm0.072$ &  \\
OGLE004756.26-725217.6 & ab & $0.592250$ & SC4\_FIV & $18.701\pm0.043$ & $18.385\pm0.037$ &  \\
OGLE004800.43-725222.7 & ab & $0.518919$ & SC4\_FIV & $18.936\pm0.047$ & $18.553\pm0.052$ &  \\
OGLE004758.51-725430.6 & ab & $0.618294$ & SC4\_FIV & $18.614\pm0.046$ & $18.319\pm0.039$ &  \\
OGLE004744.89-725530.1 & c & $0.297777$ & SC4\_FIV & $18.825\pm0.057$ & $18.855\pm0.038$ &  \\
OGLE004817.37-725536.5 & ab & $0.556467$ & SC4\_FIV & $18.690\pm0.073$ & $18.162\pm0.056$ &  \\
\hline
\multicolumn{7}{c}{2008-08-28}\\
\hline
OGLE004211.68-730329.2 & ab & $0.632047$ & SC2\_FI & $18.742\pm0.034$ & $18.443\pm0.098$ &  \\
OGLE004225.29-730349.8 & ab & $0.699868$ & SC2\_FI & $18.196\pm0.030$ & $17.898\pm0.065$ &  \\
OGLE004126.84-730355.2 & ab & $0.705041$ & SC2\_FI & $18.438\pm0.047$ & $18.256\pm0.088$ &  \\
OGLE004154.36-730539.3 & ab & $0.524564$ & SC2\_FI & $19.004\pm0.042$ & $18.836\pm0.109$ &  \\
OGLE004156.46-730641.7 & ab & $0.586298$ & SC2\_FI & $17.163\pm0.036$ & $17.260\pm0.046$ & blend \\
OGLE004817.88-731815.2 & ab & $0.558981$ & SC4\_FII & $...\pm...$ & $18.019\pm0.071$ &  \\
OGLE004744.69-731919.1 & ab & $0.598960$ & SC4\_FII & $18.470\pm0.035$ & $18.192\pm0.091$ &  \\
OGLE004805.65-732033.9 & ab & $0.523798$ & SC4\_FII & $18.892\pm0.038$ & $18.687\pm0.110$ &  \\
OGLE004724.96-731427.2 & ab & $0.614266$ & SC4\_FIII & $...\pm...$ & $...\pm...$ &  \\
OGLE004649.00-731544.5 & ab & $0.665961$ & SC4\_FIII & $18.357\pm0.034$ & $17.969\pm0.071$ &  \\
OGLE004650.36-731652.2 & ab & $0.630375$ & SC4\_FIII & $18.430\pm0.030$ & $18.237\pm0.080$ &  \\
OGLE005656.87-730850.2 & ab & $0.545403$ & SC7\_FV & $18.739\pm0.063$ & $18.346\pm0.046$ &  \\
OGLE005658.96-730850.6 & ab & $0.641613$ & SC7\_FV & $18.348\pm0.137$ & $17.271\pm0.030$ & contaminated \\
OGLE005616.96-730901.6 & ab & $0.615658$ & SC7\_FV & $18.268\pm0.032$ & $18.030\pm0.036$ &  \\
OGLE005614.52-731023.8 & ab & $0.701793$ & SC7\_FV & $16.093\pm0.014$ & $15.740\pm0.011$ & blend \\
OGLE005612.31-731222.5 & ab & $0.619898$ & SC7\_FV & $18.743\pm0.028$ & $18.517\pm0.039$ &  \\
OGLE005728.85-723454.6 & ab & $0.416258$ & SC8\_FVI & $19.615\pm0.080$ & $19.242\pm0.101$ & misidentified \\
OGLE005803.73-723602.0 & ab & $0.625510$ & SC8\_FVI & $17.385\pm0.027$ & $16.913\pm0.025$ & blend \\
OGLE005806.86-723811.7 & ab & $0.614224$ & SC8\_FVI & $18.561\pm0.033$ & $18.261\pm0.044$ &  \\
OGLE005737.26-723819.1 & ab & $0.687028$ & SC8\_FVI & $18.593\pm0.037$ & $18.301\pm0.043$ &  \\
OGLE005752.74-723901.8 & ab & $0.557047$ & SC8\_FVI & $18.883\pm0.043$ & $18.606\pm0.053$ &  \\
\hline
\multicolumn{7}{c}{2008-11-24}\\
\hline
OGLE004755.07-725141.5 & ab & $0.504549$ & SC4\_FIV & $18.454\pm0.033$ & $18.365\pm0.050$ &  \\
OGLE004805.17-725144.4 & ab & $0.509540$ & SC4\_FIV & $19.002\pm0.058$ & $18.626\pm0.058$ &  \\
OGLE004754.45-725212.1 & ab & $0.586127$ & SC4\_FIV & $18.646\pm0.055$ & $18.167\pm0.065$ &  \\
OGLE004756.26-725217.6 & ab & $0.592250$ & SC4\_FIV & $18.700\pm0.039$ & $18.446\pm0.045$ &  \\
OGLE004800.43-725222.7 & ab & $0.518919$ & SC4\_FIV & $18.746\pm0.052$ & $18.563\pm0.055$ &  \\
OGLE004758.51-725430.6 & ab & $0.618294$ & SC4\_FIV & $18.803\pm0.040$ & $18.535\pm0.055$ &  \\
OGLE004744.89-725530.1 & c & $0.297777$ & SC4\_FIV & $19.035\pm0.056$ & $18.798\pm0.067$ &  \\
OGLE004817.37-725536.5 & ab & $0.556467$ & SC4\_FIV & $18.456\pm0.050$ & $17.984\pm0.081$ &  \\
\hline
\multicolumn{7}{c}{2008-12-14}\\
\hline
OGLE005451.72-723850.4 & c & $0.277966$ & SC7\_FVII & $18.641\pm0.049$ & $18.496\pm0.058$ & contaminated \\
OGLE005527.97-724136.8 & c & $0.371404$ & SC7\_FVII & $19.081\pm0.060$ & $18.765\pm0.070$ &  \\
OGLE005621.27-730046.9 & c & $0.351118$ & SC7\_FVIII & $18.833\pm0.044$ & $18.654\pm0.103$ &  \\
OGLE005606.56-730242.1 & ab & $0.617223$ & SC7\_FVIII & $18.469\pm0.032$ & $18.167\pm0.039$ &  \\
OGLE005609.42-730323.2 & ab & $0.521781$ & SC7\_FVIII & $18.976\pm0.049$ & $18.614\pm0.055$ &  \\
\hline
\multicolumn{7}{c}{2008-12-15}\\
\hline
OGLE010409.82-724611.9 & ab & $0.644261$ & SC10\_FIX & $18.369\pm0.036$ & $18.100\pm0.060$ &  \\
OGLE010442.62-724627.5 & ab & $0.614622$ & SC10\_FIX & $18.582\pm0.045$ & $18.282\pm0.048$ &  \\
OGLE010450.99-724700.3 & ab & $0.488994$ & SC10\_FIX & $18.832\pm0.040$ & $18.567\pm0.058$ &  \\
OGLE010501.92-724932.8 & ab & $0.496103$ & SC10\_FIX & $19.258\pm0.061$ & $18.742\pm0.060$ &  \\
OGLE010448.82-724945.3 & ab & $0.650334$ & SC10\_FIX & $18.631\pm0.034$ & $18.193\pm0.043$ &  \\
OGLE010432.13-724958.4 & ab & $0.629954$ & SC10\_FIX & $18.513\pm0.035$ & $18.413\pm0.113$ &
\enddata
\label{taballstars}
\end{deluxetable}

\begin{deluxetable}{ccccccccc}
\tablewidth{0pc}
\tabletypesize{\tiny}
\tablecaption{Averaged and extinction corrected magnitudes of observed RR Lyrae stars.}
\tablehead{\colhead{Star ID [OGLE]} & \colhead{Star type} & \colhead{Period [d]} & \colhead{Field name} & \colhead{$Js$ [mag]} & \colhead{$Ks$ [mag]} }
\startdata
OGLE004211.68-730329.2 & ab & $0.632047$ & SC2\_FI & $18.663\pm0.033$ & $18.344\pm0.062$ \\
OGLE004225.29-730349.8 & ab & $0.699868$ & SC2\_FI & $18.198\pm0.028$ & $17.964\pm0.044$ \\
OGLE004126.84-730355.2 & ab & $0.705041$ & SC2\_FI & $18.485\pm0.038$ & $18.208\pm0.057$ \\
OGLE004154.36-730539.3 & ab & $0.524564$ & SC2\_FI & $18.857\pm0.036$ & $18.668\pm0.067$ \\
OGLE004817.88-731815.2 & ab & $0.558981$ & SC4\_FII & $18.356\pm0.039$ & $18.091\pm0.046$ \\
OGLE004744.69-731919.1 & ab & $0.598960$ & SC4\_FII & $18.396\pm0.036$ & $18.158\pm0.057$ \\
OGLE004805.65-732033.9 & ab & $0.523798$ & SC4\_FII & $18.735\pm0.031$ & $18.549\pm0.067$ \\
OGLE004724.96-731427.2 & ab & $0.614266$ & SC4\_FIII & $18.586\pm0.058$ & $18.385\pm0.040$ \\
OGLE004649.00-731544.5 & ab & $0.665961$ & SC4\_FIII & $18.285\pm0.031$ & $18.013\pm0.045$ \\
OGLE004650.36-731652.2 & ab & $0.630375$ & SC4\_FIII & $18.378\pm0.026$ & $18.206\pm0.048$ \\
OGLE004755.07-725141.5 & ab & $0.504549$ & SC4\_FIV & $18.340\pm0.031$ & $18.299\pm0.035$ \\
OGLE004805.17-725144.4 & ab & $0.509540$ & SC4\_FIV & $18.780\pm0.046$ & $18.536\pm0.044$ \\
OGLE004754.45-725212.1 & ab & $0.586127$ & SC4\_FIV & $18.476\pm0.051$ & $18.133\pm0.056$ \\
OGLE004756.26-725217.6 & ab & $0.592250$ & SC4\_FIV & $18.616\pm0.034$ & $18.381\pm0.034$ \\
OGLE004800.43-725222.7 & ab & $0.518919$ & SC4\_FIV & $18.756\pm0.040$ & $18.524\pm0.044$ \\
OGLE004758.51-725430.6 & ab & $0.618294$ & SC4\_FIV & $18.624\pm0.035$ & $18.393\pm0.039$ \\
OGLE004744.89-725530.1 & c & $0.297777$ & SC4\_FIV & $18.845\pm0.046$ & $18.792\pm0.044$ \\
OGLE004817.37-725536.5 & ab & $0.556467$ & SC4\_FIV & $18.488\pm0.051$ & $18.039\pm0.057$ \\
OGLE005656.87-730850.2 & ab & $0.545403$ & SC7\_FV & $18.652\pm0.063$ & $18.310\pm0.046$ \\
OGLE005616.96-730901.6 & ab & $0.615658$ & SC7\_FV & $18.181\pm0.032$ & $17.994\pm0.036$ \\
OGLE005612.31-731222.5 & ab & $0.619898$ & SC7\_FV & $18.656\pm0.028$ & $18.481\pm0.039$ \\
OGLE005806.86-723811.7 & ab & $0.614224$ & SC8\_FVI & $18.471\pm0.033$ & $18.224\pm0.044$ \\
OGLE005737.26-723819.1 & ab & $0.687028$ & SC8\_FVI & $18.503\pm0.037$ & $18.264\pm0.043$ \\
OGLE005752.74-723901.8 & ab & $0.557047$ & SC8\_FVI & $18.793\pm0.043$ & $18.569\pm0.053$ \\
OGLE005527.97-724136.8 & c & $0.371404$ & SC7\_FVII & $18.994\pm0.060$ & $18.729\pm0.070$ \\
OGLE005621.27-730046.9 & c & $0.351118$ & SC7\_FVIII & $18.746\pm0.044$ & $18.618\pm0.103$ \\
OGLE005606.56-730242.1 & ab & $0.617223$ & SC7\_FVIII & $18.382\pm0.032$ & $18.131\pm0.039$ \\
OGLE005609.42-730323.2 & ab & $0.521781$ & SC7\_FVIII & $18.889\pm0.049$ & $18.578\pm0.055$ \\
OGLE010409.82-724611.9 & ab & $0.644261$ & SC10\_FIX & $18.298\pm0.036$ & $18.071\pm0.060$ \\
OGLE010442.62-724627.5 & ab & $0.614622$ & SC10\_FIX & $18.511\pm0.045$ & $18.253\pm0.048$ \\
OGLE010450.99-724700.3 & ab & $0.488994$ & SC10\_FIX & $18.761\pm0.040$ & $18.538\pm0.058$ \\
OGLE010501.92-724932.8 & ab & $0.496103$ & SC10\_FIX & $19.187\pm0.061$ & $18.713\pm0.060$ \\
OGLE010448.82-724945.3 & ab & $0.650334$ & SC10\_FIX & $18.560\pm0.034$ & $18.164\pm0.043$ \\
OGLE010432.13-724958.4 & ab & $0.629954$ & SC10\_FIX & $18.442\pm0.035$ & $18.384\pm0.113$
\enddata
\label{tabrrlyr}
\end{deluxetable}

\begin{deluxetable}{ccccc}
\tablewidth{0pc}
\tabletypesize{\small}
\tablecaption{PL relations determined from averaged data of Table 4.}
\tablehead{\colhead{Calibration}&\colhead{$J$ slope}&\colhead{$J$ zero point}&\colhead{$K$ slope}&\colhead{$K$ zero point}}
\startdata
Free fit & $-2.772\pm0.549$ & $17.919\pm0.135$ & $-3.104\pm0.494$ & $17.598\pm0.122$ \\
Sollima et al. (2008) & $...$ & $...\pm...$ & $-2.380$ & $17.772\pm0.028$ \\
Bono et al. (2003) & $...$ & $...\pm...$ & $-2.101$ & $17.839\pm0.028$ \\
Catelan et al. (2004) & $-1.773$ & $18.159\pm0.031$ & $-2.353$ & $17.779\pm0.028$
\enddata
\label{tabpl}
\end{deluxetable}

\begin{deluxetable}{cccc}
\tablewidth{0pc}
\tabletypesize{\small}
\tablecaption{True SMC Distance Moduli determined from Different Calibrations.}
\tablehead{\colhead{Filter}&\colhead{Solima et al.}&\colhead{Bono et al.}&\colhead{Catelan et al.}\\
\colhead{}&\colhead{(2008)}&\colhead{(2003)}&\colhead{(2004)}}
\startdata
$J$ & $...\pm...$ & $...\pm...$ & $18.946\pm0.181$ \\
$K$ & $18.965\pm0.161$ & $19.002\pm0.165$ & $18.966\pm0.161$
\enddata
\label{tabdist}
\end{deluxetable}

\end{document}